\RequirePackage{fix-cm}
\documentclass[smallextended]{svjour3}       
\smartqed  
\usepackage{graphicx}
\begin{document}

\title{The  Schwarzschild  Mass in  General Relativity 
}


\author{P S Negi       
}


\institute{Department of Physics, \at
              Kumaun University, Nainital\\
              \email{psneginainital63@gmail.com}           
}

\date{Received: date / Accepted: date}

\maketitle

\begin{abstract}
The central (surface) energy-density, $E_0 (E_R)$, which appears in the expression of total static and spherical mass, $M$ (corresponding to the total radius $R$) is defined as the density measured only by one observer located at the centre (surface) in the Momentarily Co-moving Reference Frame (MCRF). Since the mass, $M$, depends only on the central (surface) density for most of the equations of state (EOSs) and/or exact analytic solutions of Einstein's field equations available in the literature, the  central (surface) density measured in the preferred frame (that is, in the MCRF) appears to be not in agreement with the coordinate invariant form of the field equations that result for the source mass, $M$. In order to overcome the use of any preferred coordinate system (the MCRF) defined for the central (surface) density in the literature, we argue for the first time that the said density may be defined in the coordinate invariant form, that is, in the form of the average density ($3M/4\pi R^3)$ of the the configuration which turns out to be independent of the radial coordinate $`r'$ and depends only on the central (surface) density of the configuration. In this connection, we further argue that the central (surface) density of the structure should be {\em independent} of the density measured on the other boundary (surface/central) because there exists no a priori relation between the radial coordinate $`r'$ and the proper distance from the centre of the sphere to its surface \cite{Ref1}. In the light of this reasoning, the various EOSs and analytic solutions of Einstein's field equations in which the central and the surface density are {\em interdependent} can not fulfill the definition of central (surface) density measured only by one observer located in the MCRF at the centre (surface) of the configuration.

\end{abstract}

\section{Motivation}

The metric for the static and spherically symmetric mass distribution can be written in the curvature coordinates as
\begin{equation}
ds^2 = e^\nu dt^2 - e^\lambda dr^2 - r^2(d\theta^2 +{\rm sin}^2\theta d\phi^2)
\end{equation}
where $G = c = 1$ (we are using geometrized units) and $\nu$ and $\lambda$ are functions of $`r'$ alone.

The Einstein's field equations for the metric given by eq.(1) can be written as
\begin{equation}
R_{ik} - (1/2)g_{ik}R = -8\pi T_{ik}
\end{equation}
where $R_{ik}$ is the Ricci tensor, $R$ is the curvature scalar, $T_{ik}$ is the stress-energy tensor and $g_{ik}$ is the metric tensor. the symbols $i$ and $k$ run from $0$ to $3$ such that $g_{ik} = 0$ if $i\neq k$. The $T^{00}$ component of the stress -energy tensor represents the energy-density, $E$, and $T^{11} = T^{22} = T^{33} = -P$ denote the isotropic pressure of the prefect fluid.

It is well known that in the correspondence limit (the case of weak field and slow motion) the field equations (eq.2) reduce to the classical Poisson Equation
\begin{equation}
\bigtriangledown^2 \phi = 4\pi \rho
\end{equation}
Recalling that we have taken $G = c = 1, \phi$ is the Newtonian gravitational potential and $\rho$ is the density of `mass'. The solution of eq.(3) for a `point' particle of mass $`M'$ is given by
\begin{equation}
 \phi = -M/r
\end{equation}
Thus the source of the gravitational field in the Newtonian Gravitation Theory (NGT) is the mass density, $\rho$, which is approximately equal to the rest-mass density, $\rho_0$. Thus it seems likely that the general relativistic generalization of this mass density should be the density of total energy including the rest-mass which is called the energy-density, $E$. However, since $E$ is the energy-density as measured only by one observer located in the MCRF which represents the $T^{00}$ component of stress-energy tensor $T$ of the perfect fluid given by eq. (2). This would introduce a preferred coordinate system (that is, MCRF) in which $T^{00}$ was evaluated. In order to avoid the use of preferred coordinate systems and to keep intact the coordinate invariant form of the field equations (eq. 2), it is argued in the literature \cite{Ref1} that the whole of the stress-energy tensor $T $ (all the components  $T^{00}$,  $T^{11}$,  $T^{22}$ and $T^{33}$) act as the source of gravitational field (or the curvature of the space-time as given in the left hand side of eq.2).

\section{Methodology}
The TOV Equations (\cite{Ref2}; \cite{Ref3}) resulting from the field equations (2) for the metric (eq.1) can be written as
\begin{equation}
P' = -(P + E)[4\pi Pr^3 + m]/r(r - 2m) 
\end{equation}
\begin{equation}
\nu'/2 = -P'/(P + E) 
\end{equation}
\begin{equation}
m'(r) = 4\pi Er^2
\end{equation}
where the prime denotes derivative with respect to $`r'$ and the mass function $m(r)$ is given by
\begin{equation}
e^{-\lambda} = 1 - [2 m(r)/r]
\end{equation}
or
\begin{equation}
 m(r) = \int_{0}^{r} 4\pi E r^2 {\rm d}r
\end{equation}
The coupled equations (5) - (7) may be solved for an assumed EOS or exact analytic relation connecting the radial coordinate $`r'$ with any of the four parameters appear in eqs.(5) - (7), subject to the following boundary conditions at the surface, $r = R$:

$P = P(R) = 0; e^{\nu(R)} = e^{-\lambda(R)} = (1 - 2M/R) = (1 - 2u)$ that ensures the continuity of mass at the surface, $m(R) = M$, which appears in the exterior Schwarzschild solution, viz. ; $e^{\nu} = e^{-\lambda} = (1 - 2M/r)$, for $r \geq R$. Thus, the total mass as measured by an external observer is given by
\begin{equation}
 m(R) = M = \int_{0}^{R} 4\pi E r^2 {\rm d}r
\end{equation}
Eq. (10) is analogous to the definition of total mass in the NGT. This analogy is termed as rather `deceptive' in the literature \cite {Ref1}, because the energy-density $E$ is measured locally whereas the integral over the volume element $4\pi r^2{\rm d}r$ is non local. The external observer measures the total mass-energy which also includes the (negative) gravitational potential energy. The possibility of the removal of  this `deception' is discussed below in the discussion.

\section{The Exact Analytic Solutions and EOSs in the Framework of Einstein's Field Equations}

There are number of EOSs and exact analytic solutions of Einstein's field equations  available in the literature \cite{Ref2}; \cite{Ref4};  \cite{Ref5};  \cite{Ref6};  \cite{Ref7} which may be categorized among three categories as given below

Category (A): The total mass, $M$,  of the configurations in this category depends only on the central density. The EOSs and analytic solutions have a positive finite density at the centre ($ e^{\lambda} = 1$) which decreases outwards (so called the `regular' solutions) and terminates at the surface together with pressure. The well known examples of EOS in this category are polytropic EOSs \cite{Ref8} and the analytic solutions in this category correspond to Tolman's VII solution with vanishing surface density  \cite{Ref2} and Buchdahl's gaseous model  \cite{Ref9}. The expression of total mass, for example, for Tolman's VII solution with vanishing surface density is given by

$$M = 8\pi E_0R^3/15$$, where $E_0$ represents the central energy-density of the structure.

Category (B): The total mass, $M$, in this category depends only on the finite (positive) value of the density at the surface where the pressure vanishes. In such configurations  the central density becomes infinity (together with pressure and $ e^{\lambda} \neq 1$ )which decreases outwards together with pressure. The total mass, $M$, however remains finite and become independent of the (infinite) central density. The examples of such exact analytic solutions are Tolman's type V and VI solutions \cite{Ref2} and the EOS in this category belongs to the well known case of non-interacting ideal Fermi gas considered in their pioneering work by Oppenheimer \& Volkoff \cite{Ref3}. The total mass $M$, for example, for Tolman's VI solution is given by the simple expression:

$$M = 4\pi E_R R^3$$, where $E_R$ represents the surface density of the structure.

Category (C): The total mass, $M$, in this category depends on the central/surface density in such a manner that the central and surface densities have become {\em interdependent}. Such structures have a positive finite density at the centre ($ e^{\lambda} = 1$) which decreases outwards with pressure and remains finite positive at the surface where the pressure vanishes.  The examples of exact analytic solutions in this category  are Tolman's IV solution \cite{Ref2} and the solutions obtained in \cite{Ref10} - \cite{Ref12} etc., whereas the examples of  EOSs in this category belongs to the case of stiffest EOS \cite{Ref13}, $\rm d P/\rm d E$ = 1 (in geometrized units) and the EOSs of strange quark matter (see, for example \cite{Ref14} and references therein). The total mass $M$, for example, in the solution \cite {Ref12} is given by the  expression:

$$M = 8X(3 + X)R/14(1 + X)^2$$
where $X = CR^2$, $C$ is a constant, and the central and surface densities are connected by the relation

$$(E_R/E_0) = (9 + 2X + X^2)/9(1 + X)^3$$

 This  interdependent feature between the central and the surface densities is clearly in contradiction with the argument mentioned in the literature that ``there is no a priori relation between the radial coordinate `$r$' and the `proper' distance between the centre to the surface of the configuration''  \cite{Ref1} which was never realized till to date. Because it clearly follows from this definition that {\em measurements of two local observers located at different boundaries (centre and surface) can not be connected simply through the radial coordinate `$r$'}. However, the author reached to the same conclusion on the basis of a different argument \cite{Ref15}. He showed that if the central and surface densities have become interdependent, the total mass, $M$, resulting from the configurations discussed under category (C) will be different from the total mass $M$ which appears in the exterior Schwarzschild solution.

\section{Discussion}

In the light of the above findings, it follows that the total mass $M$ obtained for the categories (A) and (B) solutions and EOSs depends only on the central (surface) density ( and turns out to be independent of the surface (central) density)  of the configurations. This feature is fully consistent with the definition of centre (surface) density mentioned in the literature that - it is the density measured only by one observer (MCRF) located at the centre (surface) of the configuration. The total masses $M$ of the configurations belong to category (C), on the other hand, are found to be interdependent of the central and the surface densities. This feature, therefore, can not be considered consistent with the definition of the locally measured values of the central and the surface densities which should be independent of each other as argued above in sec. 3. This fact follows from the definition mentioned above which states that ``there is no a priori relation between the radial coordinate `$r$' and the proper distance between the centre and the surface of the configuration'' \cite {Ref1}.

Yet the dependence of the total mass $M$ only on the central (surface) density further needs explanation, because the central (surface) density which is measured in the preferred coordinate system (the MCRF at the centre/surface) would become inconsistent with the coordinate invariant form of the field equations (5 - 7). Therefore, in order to keep the definition of the central (surface) density intact which follows from the literature \cite{Ref1} and to express it in the coordinate invariant form consistent with the field equations, we argue for the first time that the central (surface) density of the configuration may be defined in the coordinate invariant form of the `average (mean) density', $E_{ave}$, of the structure ($3M/4\pi R^3$) so that the average density depends only on the central (surface) density of the structure. Except the category (C), this feature is common in the structures belong to the category (A) and (B). Furthermore, the use of the `average' density in Eq. (10) of the total mass $M$ would also resolve the `deception' related with the analogy of the total mass as defined in the NGT.

The `equivalent' homogeneous density (average density $E_{ave}$) sphere of total mass $M$ and radius $R$ as measured by an external observer, corresponding to eq.(10) is given by

\begin{equation}
 m(R) = M = \int_{0}^{R} 4\pi E_{\rm ave} r^2 {\rm d}r
\end{equation}

By using the relation connecting the rest-mass density, $\rho_0$ to the energy-density, $E$ \cite{Ref16} and remembering that the rest- mass density is approximately equal to the mass density defined in the NGT \cite{Ref1}, \cite{Ref5} we get
\begin{equation}
 \rho_0 = (P + E)e^{(\nu - \nu_R)/2}
\end{equation}
Applying the boundary conditions at the surface ($P = 0$, $e^\nu = e^{\nu_R}$), for homogeneous density ($E_{ave}$) sphere eq.(12) yields

\begin{equation}
 \rho_0 = E_{ave}
\end{equation}
The substitution of eq.(13) into eq.(11) yields

\begin{equation}
 m(R) = M = \int_{0}^{R} 4\pi \rho_0 r^2 {\rm d}r
\end{equation}

which is completely analogous to the form of total mass as defined in the NGT.


\end{document}